\documentstyle[11pt,paspconf,psfig]{article}

\begin{document}

\title{Molecular lines in absorption: recent results}

\author{Fran\c{c}oise Combes}
\affil{Observatoire de Paris, DEMIRM, 61 Av. de l'Observatoire\\
F-75 014 Paris, France}

\author{Tommy Wiklind}
\affil{Onsala Space Observatory, S-43 992, Onsala, Sweden}

\begin{abstract}
Some recent results are presented about high redshift molecular absorption
lines, namely about chemical abundances of elements, and in particular of
water and molecular oxygen. Excitation temperatures of several molecules
are found lower than the cosmic background temperature at the corresponding
redshift $z=0.88582$ in PKS1830-211, and interpretations are proposed.
The radio flux monitoring of the two gravitational images of PKS1830-211
is presented over almost two years, but precise calibration is still
preventing the determination of the time-delay without ambiguity. 
The high spectral resolution of radio observations allows
to put constraints on the variation of the fine-structure constant over
a large fraction of the Hubble time.
\end{abstract}



\section{Molecules unobservable from the ground at $z=0$}

 Because of the atmospheric absorption in the H$_2$O and O$_2$ lines,
it is impossible to have a correct estimation of the abundances of
theses two molecules in the local interstellar medium.
The redshift allows us to get rid of the atmospheric absorption,
and the highest column density line-of-sights are privileged targets
to try to detect these fundamental molecules
(e.g. Combes \& Wiklind 1996).

\subsection{Water}

We have chosen the absorbing cloud in front of B0218+357, where already
optically thick lines of CO, $^{13}$CO and C$^{18}$O have been
detected (Combes \& Wiklind 1995). The optical depth of the CO(2-1)
line was derived to be 1500. The redshift is $z=0.68466$, and the
fundamental ortho transition of water at 557 GHz is redshifted
to 331 GHz into an atmospheric window.
An attempt to observe water in emission at $z=2.28$ has resulted
in a tentative detection (Encrenaz et al. 1993).

According to models, the H$_2$O/H$_2$ abundance ratio is expected 
between 10$^{-7}$ and  10$^{-5}$ (Leung et al. 1984,
Langer \& Graedel 1989). Observations of isotopic lines, such
as H$_2^{18}$O and HDO, or even H$_3$O$^+$, have confirmed these
expectations. From the ground, water maser(/thermal?) emission 
at 183 GHz has been obtained by Cernicharo et al. (1994).
However, it was thought until recently that these
abundances concerned only the neighbourhood of star-forming
regions, such as the Orion hot core, where water ice is evaporated
from grains. 
However, Cernicharo et al. (1997) detected with ISO water in
absorption at 179$\mu$ in front of SgrB2, and this revealed
that cold water was ubiquitous.

\begin{figure}
\psfig{figure=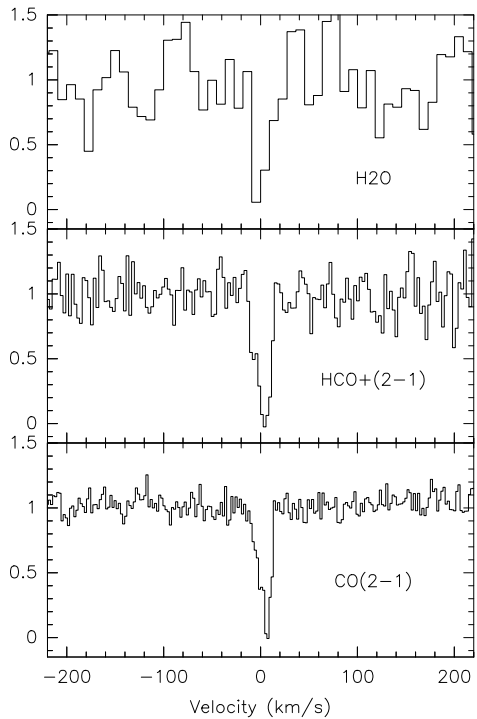,height=10.cm,angle=0}
\caption{Spectrum of ortho--water in its fundamental line
at 557\,GHz, redshifted at 331 GHz, in absorption towards B0218+357.
The line has the same width as the previously detected HCO$^+$(2--1) and
CO(2--1) lines. The velocity resolution is 9.1, 2.8 and 2.2\,km/s from
top to bottom. All spectra have been obtained with the IRAM-30m telescope,
and normalised to the continuum level completely
absorbed (33\% of the total).}
\label{fig-1}
\end{figure}

Our detection with the IRAM-30m 
of ortho-water in its fundamental line at 557 GHz
confirms this result. The line is highly optically thick, and
has about the same width as the other optically thick lines
detected in absorption in this cloud (see figure \ref{fig-1}).
If the excitation temperature was high (as in the Orion hot core),
we would have expected to detect also the excited line
at 183 GHz (redshifted at 109 GHz). An upper limit on
this line gives us an upper limit on $T_{ex}$ of 10-15K, and
an estimation of the optical depth of the 557 GHz line
of $\sim$ 40 000 (Combes \& Wiklind 1997).
This leads to an H$_2$O/H$_2$ abundance 
ratio of $10^{-5}$, in the upper range of expected values.

\subsection{Molecular oxygen}

As an element, oxygen is about twice as abundant as carbon
(O/H $\sim 8.5 \cdot 10^{-4}$), meaning that at most half
of the oxygen atoms can be used to make CO. The other half
can be found in
the form of atomic oxygen (OI), or molecules: O$_2$, OH
and H$_2$O. Since OH and H$_2$O are much less abundant than
CO, in dense molecular clouds, we expect O$_2$ to be
about as abundant as CO. Until recently, the upper limits
on the O$_2$/CO ratio was 0.1, from observations of the 
isotopic line O$^{18}$O and the direct O$_2$ in emission
in remote starbursts. The upper limits obtained through 
absorption lines are more reliable, since they involve
only one individual molecular cloud, where O$_2$
is not photo-dissociated (O$_2$ is expected
to be much less ubiquitous than CO).

We had already reported upper limits of O$_2$/CO $<$ 
1.3 $\cdot 10^{-2}$ through IRAM-30m search of
the 424 GHz and 368 GHz lines (Combes \& Wiklind 1995).
We have improved these limits by observing the
118.7 GHz line at the Kitt Peak 12m (redshifted
at 70.5 GHz), and the 56.2 GHz line at Green-Bank 43m
and Nobeyama 45m (redshifted at 33.4 GHz). 
The new limit is O$_2$/CO $<$ 2 $\cdot 10^{-3}$ at
1$\sigma$ (Combes et al. 1997).

There could be several explanations to this low oxygen abundance:

\begin{itemize}
\item  either the C/O abundance ratio in the gas phase is higher than 1.
Then all the oxygen atoms are used up in the CO molecules, the O$_2$/CO
ratio decreases exponentially with (C/O$^{-1}$). This means that the oxygen
is frozen into grains (mainly under the form of water ice or silicates).
\item the steady-state chemistry is never reached,
and because of turbulence, only time-dependent models should
apply (time-scales of less than 1.3 $\cdot 10^{5}$ yr). The
oxygen is therefore under the OI form, even in dense molecular
clouds.
 \item due to chemical bi-stability, there exists two
possible phases of the interstellar medium, the LIP and HIP
(low and high ionization phases respectively, e.g. Le Bourlot
et al. 1993, 1995). In the HIP, the O$_2$/CO abundance can be much
lower than in the LIP. It remains to be explained why the
HIP should be predominant in the ISM. Also in this model, the
abundances of other elements, such as HCN or HCO$^+$ is not
reproduced within orders of magnitude.
\end{itemize}

\section{Chemical abundances}

One of the aim of this work is to pinpoint evolution with redshift
of the molecular abundances, which could be different from the local
ones because of different element abundances, different physical environment
(temperature, densities, radiation field). Many molecules have been
detected at high redshift in absorption, HCO$^+$, HCN, HNC, CO, CS, CCH, CN, 
H$_2$O, N$_2$H$^+$, H$_2$CO, C$_3$H$_2$, HC$_3$N and isotopes; it is possible
to try now a statistical comparison of abundance ratios between these
detections and their analogues in the Milky Way. It is important to
inter-compare absorption studies (and not emission), since there are 
special biases associated to each technique (absorption traces preferentially
diffuse gas). Lucas \& Liszt (1994, 1996) have made a survey of molecular
absorptions in front of extragalactic radio sources; they find surprising
abundances, for instance HCO$^+$ one or two orders of magnitude larger than
expected, that could be explained by turbulence-induced chemistry 
(Hogerheijde et al. 1995, Falgarone et al. 1995). Curiously, they find only
diffuse clouds in their survey, but this might be due to a bias towards
high latitudes and unobscured line-of-sights. Absorptions towards the
Galactic Center (Greaves \& Nyman 1996) or in CenA
(Wiklind \& Combes 1997a) correct this bias, in view of a 
comparison with high redshift absorptions (which have sometimes high
column densities).

\begin{figure}
\psfig{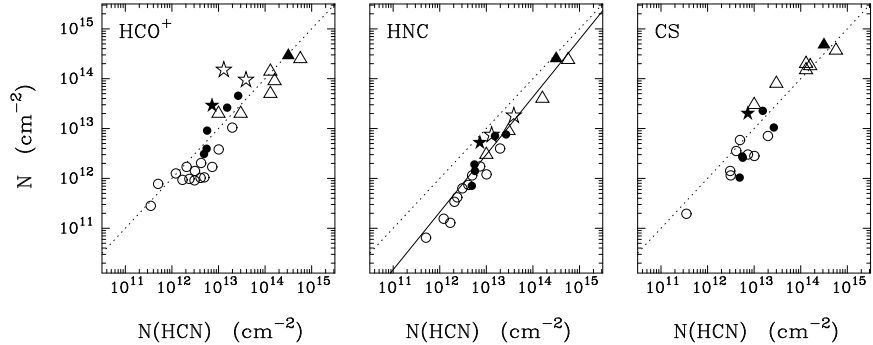}
\caption{Column density of HCN plotted versus column densitites of HCO$^+$,
HNC and CS. The open circles represent Galactic diffuse clouds (Lucas \& Liszt
1994, 1996), filled circles represent data from Cen A (Wiklind \& Combes 1997a)
and open triangles represent absorption data towards SgrB2 (Greaves \& Nyman
1996). The filled star represents our absorption data for PKS1413+135, open
stars from B3\,1504+377 at z$=$0.673 (Wiklind \& Combes 1996b) and the filled
triangle PKS1830--210 at z$=$0.886 (Wiklind \& Combes 1996a). 
The full--drawn line in $N(HCN)$ vs.
$N(HNC)$ is a linear fit to the data, showing how the $HCN/HNC$ ratio increases
with decreasing HCN column density.}
\label{fig-2}
\end{figure}

A sample of the results can be seen in figure \ref{fig-2}, where abundances
of HCO$^+$, HCN, HNC and CS are intercompared. The main striking point
is that the high-redshift abundances are perfectly compatible with
the local ones. In particular, the well known HCN/HNC variations
with physical conditions is retrieved (Irvine et al. 1987).
In fact, there are more variations from cloud to cloud
in the Milky Way than variations due to evolution.
Changes in the abundances at large distances are within the dispersion
of local variations, preventing any effects of evolution to be seen.

\section{Time-delay in PKS1830-211}

In the previous talk, Tommy has shown how the IRAM interferometer data
have confirmed that only one of the two gravitational images is
absorbed by molecular clouds at the main velocity (V=0 or $z=0.88582$,
Wiklind \& Combes 1997c).
The SW image is entirely covered at V=0, which is also compatible with
the BIMA data (Frye et al. 1997), while the NE image is absorbed by
clouds at V $\sim$ -150km/s (may be only partially). Since the
absorption at V=0 is optically thick, its depth is a good indicator 
of the continuum level of the SW image, while the total continuum
detected with a single dish (without resolving the two images) is
a measure of the sum of the NE+SW continuum levels. By a single
observation, it is therefore possible to derive
separately the continuum of both images. Through a monitoring of
the source in its HCO$^+$(2-1) absorption, we can derive the light
curves of the NE and SW images, and try to determine the time-delay.

\begin{figure}
\psfig{figure=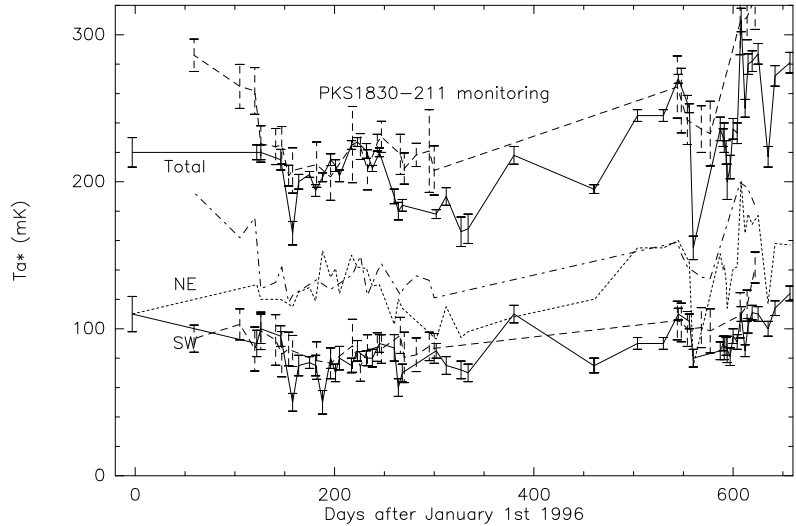,width=12.cm,angle=-90}
\caption{Results of the weekly monitoring of the quasar
PKS 1830-211 in the HCO$^+$(2-1) line at $z=0.88582$. The full and dashed lines
represent observations done at the IRAM-30m and SEST-15m telescopes
respectively. From bottom to top the curves are the measure of the continuum
level successively of image B, A and total. An intrinsic level increase appears
from 1996 to 1997, in the two images.}
\label{fig-3}
\end{figure}

We have carried out a weekly monitoring since the beginning of 1996
with the IRAM-30m and SEST-15m telescopes, the results are plotted
in figure \ref{fig-3}. The SEST data have been normalised to be
compared with the IRAM ones, both are pretty compatible.
However, it is difficult to derive precisely the time-delay, since
the intrinsic variations of the quasar have not been of high amplitude
1996-7, and the atmospheric calibrations introduce unwanted
noise in the light curves. The expected time-delay is of the order
of a few weeks.
Another caveat, in view of determining the Hubble constant,
is that a second lens at $z=.19$ could add some 
amplification or shearing effect (Lovell et al. 1996).

\section{Cosmic background temperature}

When the absorption occurs in relatively diffuse gas, where the density
is not enough to excite the rotational ladder of the molecules, the
excitation temperature is close to the cosmic background temperature
$T_{bg}$, and could be a way to check its variation with redshift.
This is the case of the gas absorbed in front of PKS1830-211, where 
$T_{ex} \sim T_{bg}$ for most of the molecules. The measurement of
$T_{ex}$ requires the detection of two nearby transitions. When the 
lower ones is optically thick, only an upper limit can be derived
for $T_{ex}$. Ideally, the two transitions should be optically thin, but 
then the higher one is very weak, and long integration times are
required.

\begin{figure}
\psfig{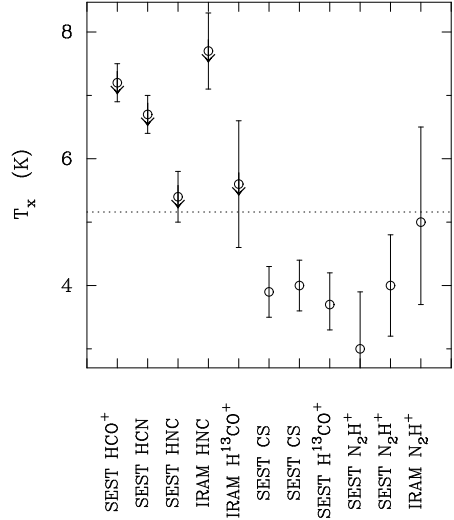}
\caption{ Measure of the excitation temperatures for several molecules shown
in abcissa, from two of their rotational transitions. When the lower 
transition is optically thick, only an upper limit is derived. The horizontal 
dash line is the cosmic background temperature expected from the big-bang
at the redshift of the absorber for PKS1830-211, i.e. $z=0.88582$.}
\label{fig-4}
\end{figure}

The results obtained with the SEST-15m and IRAM-30m are plotted in
figure \ref{fig-4}. Surprisingly, the bulk of measurements
points towards an excitation temperature lower than the background
temperature at $z=0.88582$, i.e. $T_{bg}$= 5.20K. This could be
due to a non-LTE excitation. In fact, the excitation temperature 
is not the same for each couple of levels considered. If the effect of
collisional excitations could be neglected, the levels would be in
radiative equilibrium with the black body at $T_{bg}$ (the time-scale
for this, of the order of A$^{-1}$, is less than a few years years 
for all molecules). 
But collisions tend to excite the lower levels at a higher temperature
(since $T_{kin} > T_{bg}$). The competition between the two processes 
is traced by the C/A ratio (collisions versus spontaneous rates), which 
is non-negligible only for the first level, at low density. This will
populate the $J=1$ level a little bit higher with respect to the $J=0$
than expected from radiative equilibrium at $T_{bg}$. The consequence
is that the $T_{ex}$(1-0) will be higher than $T_{bg}$ but $T_{ex}$(2-1)
will be lower than $T_{bg}$. Since the $T_{ex}$ measured in figure
\ref{fig-4} do not involve the fundamental levels (the corresponding
transitions, in the cm domain, would be optically thick), this could
be the explanation. A detailed non-LTE model should be built to
confirm this hypothesis.

\section{Variation of fine-structure constant}

 The high spectral resolution of heterodyne techniques, the narrowness of
absorption lines and their high redshift make these measurements 
favorable to try to refine the constraints on the variation of
coupling constants with cosmic time, variations that are
predicted by for instance the Kaluza-Klein and superstring theories.
By comparing the HI 21cm line (Carilli et al. 1992, 1993) with rotational
molecular lines, one can constrain the variations of
$\alpha^2 g_p$, $\alpha$ being the fine-structure constant, and
$g_p$ the proton gyromagnetic ratio. Also, by intercomparison of
rotational lines from different molecules, one can test the invariance
of the nucleon mass $m_p$, since the frequencies are affected by centrifugal
stretching.

Recent works on these lines have considerably improved the previous
limits (e.g. Potekhin \& Varshalovich 1994). By comparing various
optical/UV lines (of H$_2$, HI, CI, SiIV) in absorption in front of quasars, 
Cowie \& Songaila (1995) constrained the variation of $\alpha^2 g_p m_e/m_p$
to 4 $\cdot 10^{-15}$/yr. Varshalovich et al. (1996) from radio lines
come to a limit of variation of $\alpha^2 g_p$ of 8 $\cdot 10^{-15}$/yr.
Drinkwater et al. (1997) by a more careful analysis of the same data
conclude to 5 $\cdot 10^{-16}$/yr. We have also derived a limit
from PKS1413+135 and PKS1830-211 data of $\Delta z/(1+z) < 10^{-5}$,
which yield a corresponding limit of 2 $\cdot 10^{-16}$/yr
(Wiklind \& Combes 1997b).
However, geophysical constraints are in fact superior to all
astrophysical ones. Damour et al. (1997) have recently come up
with a limit of 5 $\cdot 10^{-17}$/yr on $\alpha$ from the natural fission 
reactors which operated about  2 $\cdot 10^{9}$yr ago at Oklo (Gabon).
 These results were obtained through analysis of the neutron capture
cross section of Samarium, in the Oklo uranium mine.

Notice that we have reached an intrinsic maximum of precision
with the astrophysical technique, since the limitation comes
from the hypothesis that the various lines compared come from
the same material, at exactly the same Doppler velocity along
the line-of-sight. This hypothesis is obviously wrong when
comparing HI and molecular lines; it is also wrong while intercomparing
molecules, or even within lines of the same molecule,
since opacity depends on excitation conditions which vary along the
line of sight for each transition.


\end{document}